\documentclass[prc,aps,floats,floatfix,twocolumn,showpacs,nofootinbib,%
superscriptaddress]{revtex4}

\usepackage{amsmath}
\usepackage{amssymb}
\usepackage{amsfonts}
\usepackage{graphicx}
\usepackage{tabularx}
\usepackage{multirow}
\usepackage{lscape}
\usepackage{rotating}

\usepackage{pstricks}

\begin{document}

\title{Pairing properties and specific heat of the inner crust of a neutron star.}


\author{A. Pastore}
\affiliation{Institut d'Astronomie et d'Astrophysique, Code Postal 226, Universit\'e Libre de Bruxelles, B-1050 Brussels, Belgium}


\begin{abstract}
We investigate the pairing properties at finite temperature of the Wigner-Seitz cells in the inner crust of a neutron star obtained with the recent Brussels-Montreal Skyrme functional BSk21. In particular we analyze the phenomena of persistence and reentrance of  pairing correlations and their impact on the specific heat  in the low-density region of the inner crust. 
\end{abstract}


\pacs{ 26.60.Gj, 21.60.Jz, 21.30.-x}
 
\date{\today}


\maketitle

%
\section{Introduction}
\label{sect:intro}

The inner crust of a neutron star (NS), although it represents a very small fraction of the total mass of the star, plays a crucial role in a variety of phenomena~\cite{hae07,cha08} and in particular on its thermalization process. It is thus very important to study the thermal properties of the different constituent of the crust~\cite{gne01,for10,pag04}.
According to standard models, the inner crust of NS is formed by a lattice of neutron rich nuclei immersed in a sea of free neutrons and ultra-relativistic electrons~\cite{cha08} and characterized by  baryonic densities ranging from $\rho_{b}\approx5\times10^{11}\text{ g/cm}^{3} $ to $\rho_{b}\approx10^{14}\text{ g/cm}^{3} $.
A very convenient model used to describe this region of the NS is based on the Wigner-Seitz (WS) approximation~\cite{wig33}. Following the pioneering article of Negele and Vautherin~\cite{Negele1973}, we consider  spherical cells of radius $R_{WS}$ centered on each cluster in such a way to cover the entire volume of the crust. The WS cells are non-interacting and electrically neutral. In  Ref.~\cite{cha07} Chamel \emph{et al.} have investigated the validity of this approximation, showing that it can be considered as a reliable model up to baryonic densities   $\rho_{b}\approx8\times10^{13}\text{ g/cm}^{3}$. 

Comparing cooling calculations with available estimates for the surface temperatures of NS, several groups have estimated the presence of a superfluid phase~\cite{page1992,pizzochero1991,lattimer1994}. The presence of pairing correlations within the inner crust directly affects its thermal properties. It is thus very important to perform microscopic calculations which could then be used for successive astrophysical studies.
In a recent article, Fortin \emph{et al.}~\cite{for10}, performing  Finite Temperature Hartree-Fock-Bogoliubov (FT-HFB) calculations in WS cells, have shown that the thermal evolution of pairing correlations is quite different for  low  and high density WS cells.
By inspecting the neutron specific heat of the low-density WS cells given in Ref.~~\cite{Negele1973}, they observed the presence of two discontinuities corresponding to the disappearance of superfluidity in two regions of the cell: the gas and the cluster. The result has been later confirmed in Refs.~\cite{pas12,pas14}. Within the standard BCS theory~\cite{Book:Ring1980,DeGennes1999}, one can see that the superfluid phase disappears beyond a certain value of the temperature of the system,  $T_{c}$, since Cooper pairs are broken due to thermal fluctuations. 
For an homogenous system, the critical temperature can be related to the pairing gap at zero temperature, $\Delta_{T=0}$, as~\cite{abr75}

\begin{equation}\label{Tcrit:bcs}
k_{B}T_{c}\approx0.57\Delta_{T=0}\,,
\end{equation}

being $k_{B}$ the Boltzmann constant.
\noindent This result has been also validated to be a good approximation in inhomogeneous systems as finite nuclei and high-density WS cells~\cite{kha06,chamel10,niu13}.
In the region between the outer and the inner crust, the evolution of the pairing gap with the temperature is much richer then predicted by the simple BCS theory.
In Ref.\cite{mar12}, Margueron and Khan have shown how the coupling between bound and continuum states plays an important role in the suppression and persistence of pairing correlations. In the present article, we will continue the analysis we have started in Ref.~\cite{pas13marg}, by studying the effect of the coupling with the continuum states on the specific heat of the WS cell.
The article is organized as follows: in Sec.\ref{sec:T0}, we study the pairing correlations at zero temperature for some selected nuclei and WS cells, while  in Sec.\ref{pair:Tfinita} we consider the thermal properties of the inner crust at low density. Finally we give our conclusions in Sec.~\ref{sec:conclusion}.

%
%
\section{Pairing properties at T=0}\label{sec:T0}

To investigate the thermal properties of the inner crust we solved the FT-HFB equations~\cite{Goodman1984} in a spherical box of radius $R_{WS}$ and using the Dirichlet-Neumann mixed boundary conditions~\cite{Negele1973}. 
All the relevant numerical methods on the solution of these equations as well on their accuracy have been already presented in Refs.~\cite{pas11,pas12,pas13marg} and we thus omit to repeat them here. To have a simpler notation, we adopt a system of natural units where $\hbar=c=k_{B}=1$.

The FT-HFB equations are solved using  the recent Brussels-Montreal Skyrme functional BSk21~\cite{gor09}. This functional is well suited for astrophysical calculations since it has been  built to reproduce with high accuracy all experimentally known masses of atomic nuclei ($\approx2000$) with a root-mean-square deviation of $\sigma= 0.58$ MeV.
Moreover, the functional has been also fitted to reproduce the Equation of State (EoS) of Li and Schulze~\cite{li08} in pure neutron matter (PNM). This is a very important feature for calculations of systems with strong isospin asymmetry.
For the pairing channel, the BSk21 has been equipped by a zero range interaction whose parameters are constrained to reproduce the $^{1}$S$_{0}$ gap in PNM and symmetric nuclear matter (SNM) obtained by realistic calculations~\cite{cao06}. See also Ref~\cite{cha08b} for more details.

Thanks to all these features, BSk21 can be considered a good functional to be used for astrophysical calculations.

\subsection{Inner crust}
 
The chemical composition of the inner crust of the NS has been studied by Pearson \emph{et al.} ~\cite{duc12,pea14} using the the BSk21 Skyrme-functional.
They have performed the minimization of the total energy of the WS cell at $\beta$ equilibrium and at zero temperature using extended Thomas-Fermi plus Strutinsky integral method (ETFSI)~\cite{onsi97}. They have found that the most favorable configuration all along the inner crust is made by $Zr$ isotopes. 

%
 
The chemical composition of the inner crust has been investigated also by other groups using different methods and different functionals~\cite{bal06,onsi08,Negele1973,gri11}, showing a certain preference for nuclei close to $Z=40$ and to $Z=50$. Since the energy minima are relatively close to one another, the different approximations adopted during the calculations could play a non-negligible role, as for example the treatment of states in the continuum~\cite{mar07} or the effect of superfluidity~\cite{pea14B}. 
A systematic comparison among the different models would be thus important for a better insight of the physics of this system. This analysis go beyond the scope of the present article. For consistency, we will adopt the chemical composition obtained with ETFSI for BSk21.

\begin{figure}
\begin{center}
\includegraphics[width=0.42\textwidth,angle=-90]{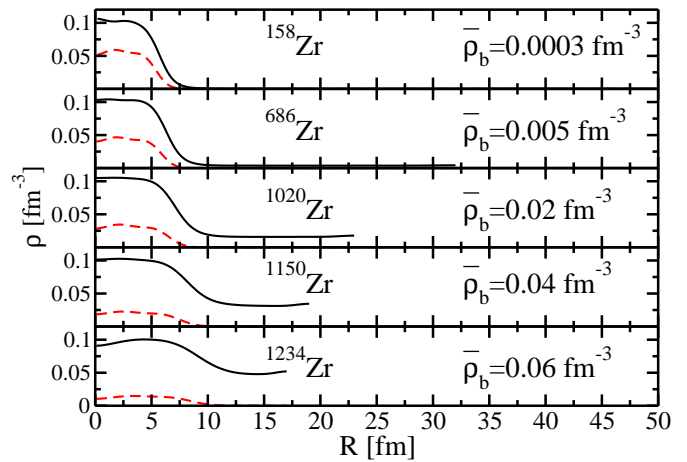}
\end{center}
\caption{(Color online) Neutron (solid line) and proton (dashed line) densities obtained with  the WS configurations given in Ref.~\cite{duc12}.}
\label{densCrust}
\end{figure}

\begin{figure}
\begin{center}
\includegraphics[width=0.42\textwidth,angle=-90]{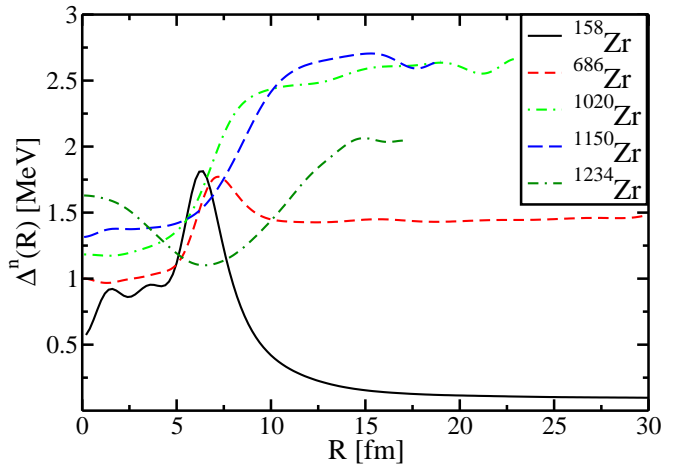}
\end{center}
\caption{(Color online) Neutron pairing field for different WS cells calculated at zero temperature.}
\label{pairCrust}
\end{figure}

\begin{figure}
\begin{center}
\includegraphics[width=0.42\textwidth,angle=-90]{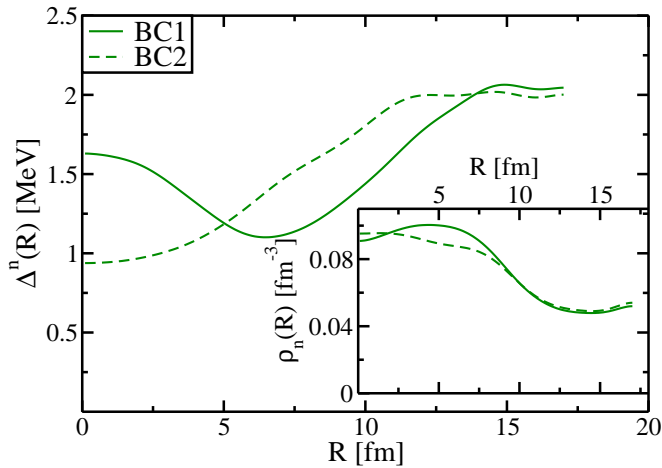}
\end{center}
\caption{(Color online) Neutron pairing field for $^{1234}$Zr using two different sets of boundary conditions as defined in Ref.~\cite{bal06}. In the insert, we show the neutron density $\rho_{n}(R)$.}
\label{pair1234zr}
\end{figure}

In Fig.\ref{densCrust}, we show the densities of  the WS cell at different average baryonic densities, $\bar{\rho}_{b}$, as obtained from Ref.~\cite{duc12} and calculated using the complete BSk21 functional.
This result can be directly compared with Fig.6 of Ref.~\cite{duc12}. Our results are in good agreement with the ETFSI one up to $\bar{\rho}_{b}=0.04\text{ fm}^{-3}$, beyond this value we notice that our  results are affected by spurious shell effects in the external neutron gas. These effects are related to the discretization of the continuum in the box~\cite{bal06,cha07,mar07}.
Compared to the EFTSI method, we also observe the presence of small ripples in the cluster region. To some extent these are artifacts of the mean-field approach used here and they are expected to be washed out by the inclusions of correlations~\cite{yao12}%
In Ref.~\cite{duc12}, there is an additional WS cell at $\bar{\rho}_{b}=0.08$ fm$^{-3}$ which we have omitted here.  For this case, our  calculations converges toward a solution where a large fraction of protons sits at the edge of the cell. This result is very sensitive to the choice of the boundary conditions  and on the initial guess on mean field potential used to solve FT-HFB equations. We refer to Ref.~\cite{pas11} for a more detailed discussion. For such a reason we discarded this cell from our analysis.
In Fig.\ref{pairCrust}, we show the neutron pairing field, $\Delta^{n}(R)$ for the WS cells  shown in Fig.\ref{densCrust}. In the high density region, $\bar{\rho}_{b}\ge0.02$ fm$^{-3}$, the main contribution to pairing correlations comes form the external gas and  the cluster acts like an impurity which decreases the gap compared to the homogeneous case~\cite{pizzo02}. This is a general behavior which does not depend on the composition of the cluster~\cite{pas11}.
For the low density region, $^{158}$Zr, the situation is the opposite: the neutron pairing field is rather weak in the external neutron gas, while it is mainly concentrated in the cluster region. In this case $\Delta^{n}(R)$ strongly depends on the properties of the cluster and in particular on its shell structure. For the WS $^{686}$Zr, we observe an intermediate behavior: it is still possible to observe a small peak at the surface of the cluster, similarly to $^{158}$Zr. 
The pairing field  for $^{1234}$Zr is quite different from the $^{1150}$Zr although they differ by only 84 neutrons. 
To clarify this very anomalous behavior, we have repeated our calculations by changing the choice of the boundary conditions at the edge of the box as discussed in  Ref.~\cite{bal06}. The  Dirichlet-Neumann mixed boundary conditions~\cite{Negele1973} can be obtained in two ways: (i) even-parity wave functions  and first derivative of odd-parity wave-functions vanish at $R=R_{WS}$ (BC1), (ii) the other way round (BC2). Since the choice of the boundary conditions is arbitrary, the result should not depend on it.
In Fig.\ref{pair1234zr}, we compare the neutron pairing field obtained with the sets BC1 and BC2.
We clearly notice, that the pairing field strongly depends on this choice thus showing that the method used to solve HFB calculations is no more adapted in this case. This result is  consistent with previous findings of Ref.~\cite{cha07} concerning the validity of the WS approximation.
We thus retire this WS cell from our successive analysis.

\subsection{Neutron drip-line}

From the ETFSI  calculations  done in Ref.\cite{duc12}, the most energetically favorable configuration for the WS cells in the crust is the one with clusters made by Zirconium isotopes. Anyhow, it is worth noticing that  within a difference of few KeV of energy per particle, we can find other proton numbers as for example $Z=50$.
The inclusion of thermal effects would probably lead to a mixed configuration.
For such a reason, we will consider in our study of the interface between outer and inner crust two isotopic chains: Zirconium and Tin.

According to the general result shown in Refs.~\cite{sch11,pas12PRA}, when a quantal system drips out from a very  small to a very large potential (container), pairing correlations at the Fermi surface are suppressed at the drip point. In this case, due to the very large number of atoms, shell effects are washed away.
By reducing the number of particles to few hundreds as in the nuclear case, the result can be affected by the specific underlying single particle structure, and in particular to the position of low-energy resonant states~\cite{kru01} in the single particle spectrum~\cite{pas13marg,mar12}. 
We can build pairing correlations between bound and resonant  states  unless there is a large shell gap between them (compared to the strength of the pairing gap). The $Zr$ and $Sn$ have been  identified as good examples for these two different behaviors. We have thus performed systematic HFB calculations  at zero temperature for these two isotopic chains by placing each nucleus at the center of a spherical box of $R_{WS}=50$ fm radius. 
\begin{figure}
\begin{center}
\includegraphics[width=0.42\textwidth,angle=-90]{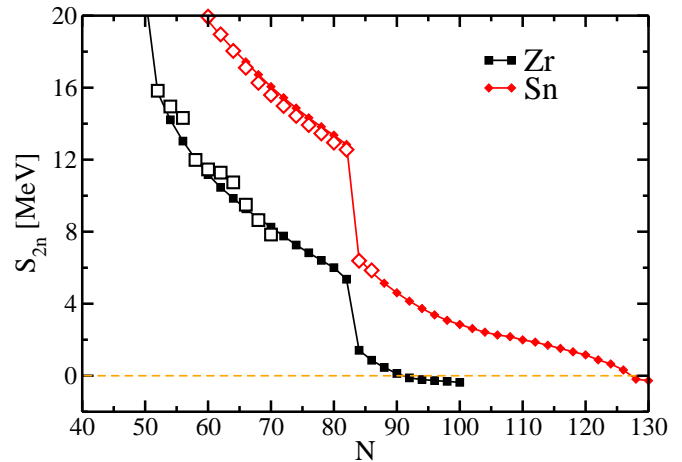}
\end{center}
\caption{(Color online) Two neutron separation energies $S_{2n}$ for Zr and Sn chains. The full symbols represent the HFB calculations, while the open symbols are the experimental value taken from  Ref.~\cite{aud12}. }
\label{S2n}
\end{figure}

To identify the position of the two-neutron drip-line, we have analyzed the two-neutron separation energy $S_{2n}$. The result is represented in Fig.\ref{S2n}.  The two drip-line nuclei for BSk21 are $^{130}$Zr and $^{176}$Sn. On the same figure, the symbols represent the experimental data point taken from Ref.~\cite{aud12}. The position of the two neutron drip line depends clearly on the choice of the interaction. We refer to Ref.~\cite{erl12} for a more systematic analysis.

\begin{figure}[!h]
\begin{center}
\includegraphics[width=0.42\textwidth,angle=-90]{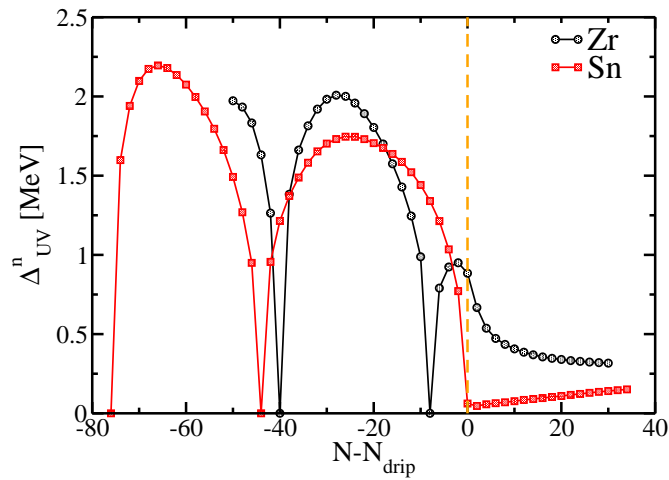}
\end{center}
\caption{(Color online) Average neutron pairing gaps as defined in Eq.\ref{eq:avgap} for $Zr$ and $Sn$ isotopic chains. See text for details.}
\label{gaps}
\end{figure}

To quantify the presence of pairing correlations, we defined the average pairing gap as
\begin{eqnarray}\label{eq:avgap}
\Delta^{q}_{UV}=\frac{\int d^{3}r \kappa^{q}(r) \Delta^{q}(r)}{ \int d^{3}r \kappa^{q}(r)  }\,,
\end{eqnarray}
\noindent being $\kappa^{q}(r)$ the pairing tensor for neutrons ($q=n$) and protons ($q=p$) respectively~\cite{Book:Ring1980}. 
This definition is well adapted to describe pairing properties of overflowing systems, since it averages pairing correlations on several states around the Fermi energy, $\varepsilon_{F}^{q}$, which belong to the gas and to the cluster.
In Fig.\ref{gaps}, we show the evolution of the averaged neutron gap $\Delta^{n}_{UV}$ as a function of the neutron number. To put in evidence the behavior at the drip-line, we have rescaled the $x$-axis by subtracting the number of neutron at the drip line $N_{drip}$ as done in Ref.~\cite{mar12}. %

We notice that for $Sn$ isotopes, once we pass the drip-line there is a strong reduction of the pairing gap. Beyond the drip-line the pairing correlations are essentially formed in the free neutron gas, while for $Zr$ isotopes the  gap $\Delta^{n}_{UV}$ does not go to zero beyond the drip-line and it stays at a value which is bigger than the one obtained by considering only the contribution of the free neutron gas  (see also Fig.5 of Ref.~\cite{pas13marg}).
We can understand this difference by looking at the single particle structure: in Tab.\ref{tab:cano}, we show the most relevant  single particle states, $\varepsilon_{lj}$, for $^{130}Zr$, which are obtained as the eigensolutions of Hartree-Fock (HF) Hamiltonian after the final convergence of the  HFB calculation~\cite{zha12}. $(l,j)$ stand for the orbital and total angular momentum of the particle. We observe the presence of  two resonant states $p_{1/2},f_{5/2}$ close to threshold and one very loosely bound state $p_{3/2}$. Due to the specific spatial extension of their wave-function, they have a quite strong overlap with the wave-functions of bound states in the last open shell, thus leading to non-zero matrix elements of the gap.
The situation for $Sn$ isotopes is quite different since $^{176}$Sn represents the neutron shell closure at $N=126$, thus the last bound  single-particle level  is $\varepsilon_{i_{13/2}}=-1.67$ MeV, thus representing a quite strong shell gap that prevents from the possibility to couple with  continuum states. 
These results  are in good agreement with previous calculations done with different pairing functionals and different Skyrme functionals~\cite{pas13marg}.
The inclusion of thermal fluctuations can change this picture since they modify both the occupation of the levels and the shell structure. We discuss this effect  in Sec.\ref{pair:Tfinita}.

\begin{table}
\setlength{\tabcolsep}{.15in}
\renewcommand{\arraystretch}{1.6}
\begin{center}
\begin{tabular}{cccc}
\hline
\hline
$\varepsilon_{lj}$ [MeV]   & $\Gamma_{lj}$ [MeV]&$l$ &$j$ \\
\hline
   3.20     & 0.05  & 5           & 9/2 \\
   1.78     &   0.5  &3          & 5/2\\
  0.48    &   3.5     &1           &1/2\\
  -0.03&        -   & 1       &    3/2\\
 -0.27  &       -    &3         &  7/2\\
  -4.39 &       -     &    5        &  11/2\\
\hline
\hline
\end{tabular}
\caption{Single-neutron energies  for $^{130}$Zr obtained with the BSk21 functional. $\Gamma_{lj}$ is the width of the resonant states. The neutron chemical potential is $\mu^{n}_{F}=-0.06$ MeV. See text for details.}
\label{tab:cano}
\end{center}
\end{table}

\section{Pairing properties at finite temperature}\label{pair:Tfinita}

\subsection{Pairing field}
\begin{figure}
\begin{center}
\includegraphics[width=0.42\textwidth,angle=-90]{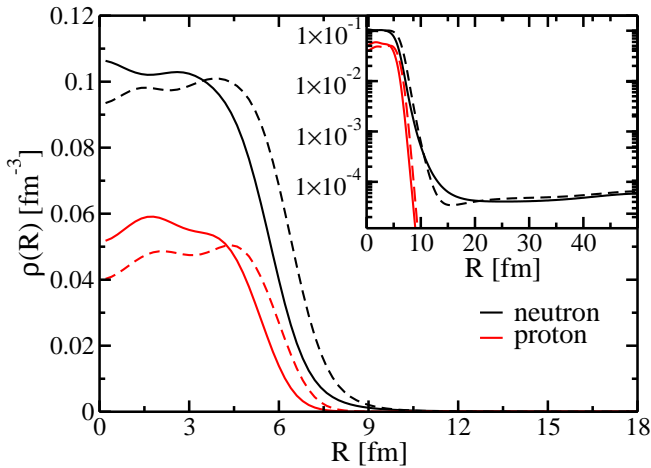}
\end{center}
\caption{(Color online) Neutron and proton density for the $^{158}$Zr (solid lines) and $^{204}$Sn (dashed lines) at zero temperature.}
\label{158zrdensity}
\end{figure}

In this section, we analyze the impact of thermal effects on pairing correlations.
For the following discussion, it is  interesting to consider two WS cells: namely  $^{158}$Zr, which is the result of a complete minimization with ETFSI method and $^{204}$Sn. 
The latter has been constructed by taking $Z=50$ protons and adjusting the number of neutrons to have the same value of the density of the external neutron gas as in  $^{158}$Zr, $i.e.$ $\rho_{gas}\approx5\times10^{-5}$~fm$^{-3}$.
In Fig.\ref{158zrdensity}, we show the density profile $\rho_{q=n,p}$ of neutrons and protons for these two WS cells. 

\begin{figure}
\begin{center}
\includegraphics[width=0.42\textwidth,angle=-90]{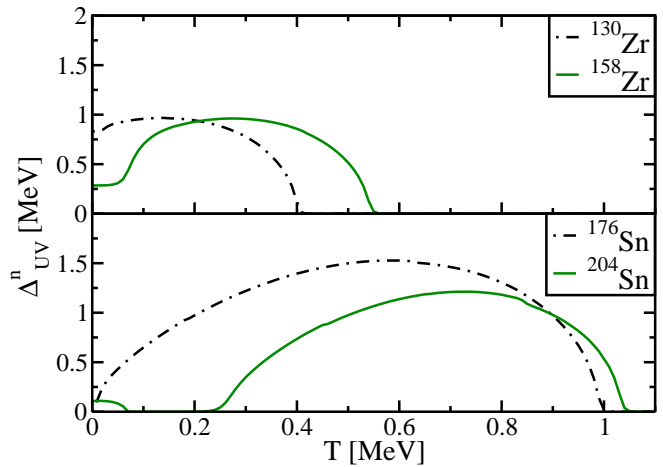}
\end{center}
\caption{(Color online) Average neutron pairing gap $\Delta^{n}_{UV}$ as a function of the temperature for different systems. See text for details. }
\label{gapT}
\end{figure}

In Fig.\ref{gapT}, we show the evolution with the temperature of the average neutron pairing gap $\Delta^{n}_{UV}$ as defined in Eq.\ref{eq:avgap} for these two WS cells.
At $T=0$, the cell $^{204}$Sn is superfluid, although the neutron pairing gap is relatively small $\Delta^{n}_{UV}=0.1$ MeV. By increasing the temperature, we observe a phase transition at $T=60$ KeV, where the neutron pairing gap drops to zero. 
We still increase the temperature and at $T=0.24$ MeV, we observe that the neutron pairing gap starts to increase again.
When the temperature is larger than the critical value $T^{2}_{c,n}=1.04$ MeV, pairing correlations are completely suppressed. 

For $^{158}$Zr, the situation is quite different: at $T=0$ the neutron pairing gap is $\Delta^{n}_{UV}=0.28$ MeV, by increasing the temperature, the pairing gap increases and it reaches its maximum at $T=0.30$ MeV and $\Delta^{n}_{UV}=0.96$ MeV and then it decreases again until it disappears beyond a critical value $T_{c,2}^{n}=0.55$ MeV.
Similar results have been also presented in Ref.~\cite{mar12} for other Skyrme functionals.

In Fig.\ref{gapT}, we also show the evolution with the temperature of the neutron pairing gap for the two nuclei at the drip-line. We recall that the FT-HFB theory is not adapted to describe the thermal properties of isolated nuclei~\cite{gam12,gam13}. 

For $^{130}$Zr, the critical temperature is $T_{c}^{n}=0.40$ MeV, while for $^{158}$Zr the critical temperature is $T_{c,2}^{n}=0.55$ MeV. By adding 28 neutrons which form a very diluted gas, we observe  an increase of $\approx30$\% of the value of the critical temperature, meaning that pairing correlations are stronger in this case.
The difference between the critical temperatures between  $^{176}$Sn and $^{204}$Sn is much smaller, in fact  $T_{c}^{n}=1.0$ MeV for $^{176}$Sn. We thus observe a difference of only  $\approx$4\% between the two critical temperatures. The number of neutrons  forming the gas is the same in the two cases. 

\begin{figure}[!h]
\begin{center}
\includegraphics[width=0.42\textwidth,angle=-90]{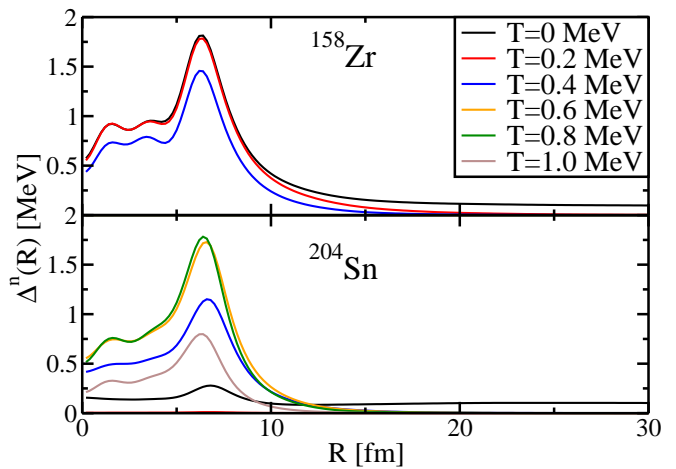}
\end{center}
\caption{(Color online) Neutron pairing field for $^{158}$Zr (upper panel) and $^{204}$Sn (lower panel) calculated at different values of the temperature of the system.  }
\label{158pair}
\end{figure}

In Fig.\ref{158pair}, we show the neutron pairing field for the two WS cells ($^{158}$Zr,$^{204}$Sn)  at different values of the temperature. Contrary to  $\Delta^{n}_{UV}$  which is an average quantity, $\Delta^{n}(R)$ gives us more information about the different components of the system, namely the cluster and the gas.
For $T=0$, we observe that the pairing field of the cell  $^{204}$Sn is rather uniform in both the gas and cluster region. At $T=0.2$MeV it drops to zero, than it starts to increase again in  the temperature interval $T\in[0.23,1.04]$ MeV, but only in the cluster region.

\begin{figure}
\begin{center}
\includegraphics[width=0.42\textwidth,angle=-90]{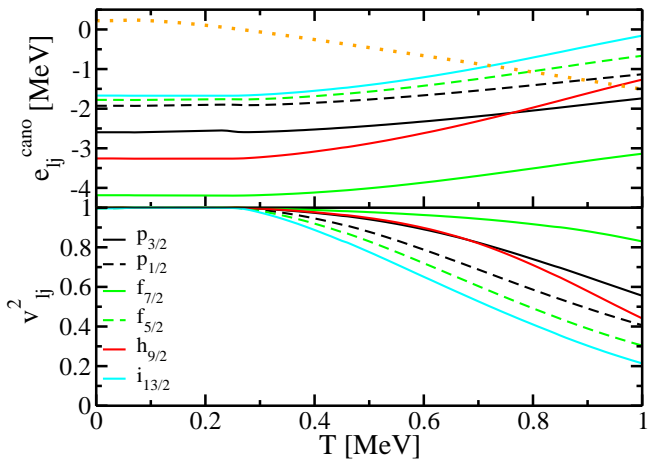}
\end{center}
\caption{(Color online)   Evolution of canonical neutron states (upper panel) and their occupation probability (lower panel) as a function of the temperature for $^{204}$Sn. The dotted line represents the evolution of the neutron chemical potential $\mu^{n}_{F}$.}
\label{levelsn204}
\end{figure}

The presence of the temperature modifies the occupation probabilities of the last major shell creating $holes$, thus allowing the formation of pairing correlations between these states.
In Fig.\ref{levelsn204}, we show the occupation probabilities of the canonical neutron states, $v^{2}_{lj}$, of the last major shell for $^{204}$Sn. At $T=0$ all the states are fully occupied and thus they do not contribute to superfluidity, as seen in Fig.\ref{158pair} pairing correlations arise essentially from gas states.
At $T=0.23$ MeV the levels starts to be unoccupied and thus we can use them to build pairing correlations, consistently with the reappearance of the pairing field in Fig.\ref{158pair}.
We also notice that not only the occupations probabilities are modified by the temperature, but also the canonical energies $e^{cano}_{lj}$ which are close to threshold. Although the change in occupation and energy shift are strongly correlated, we conclude that the main effect on pairing reentrance comes from the formation of $holes$ in this shell.

In the case of $^{158}$Zr, we observe from Fig.\ref{158pair} that the pairing field is suppressed in the gas at $T_{c,1}^{n}=60$ KeV, while it persists in the cluster, until at $T_{c,2}^{n}=0.55$ MeV the entire system is no more superfluid.
As previously discussed for the $^{130}$Zr case, the superfluidity mainly arises from scattering of pairs between loosely bound states and low-energy resonant states. 
These states are not fully occupied as in $^{204}$Sn and thus we have no reentrance phenomena. Adding temperature effects, we change the occupation probabilities of these states. This can be easily observed by looking at the canonical basis representation.
Within the interval $T\in[0,0.2]$ MeV, the number of neutrons occupying these states is not modified, at higher temperatures the neutron occupy other scattering states at higher energy and we suddenly observe a reduction of the pairing field, see Fig.\ref{158pair}.
At  $T_{c,2}^{n}=0.55$, pairing correlations are suppressed and the occupation probability of the low-lying resonant states is strongly reduced compared to the $T=0$ case. The different behavior of these two systems at finite temperature indicates the necessity of performing an analysis of the chemical composition of the inner crust at different values of the temperature. In fact the energy differences between the different  minima found in Refs.~\cite{duc12,pea14} could be strongly modified by temperature effects.

\subsection{Specific heat}

We now discuss the main features of the specific heat for the different components of the WS cells. 
The electrons in the WS cell can be treated as a uniform ultra-relativistic gas and their specific heat obtained trough the standard linear approximation~\cite{for10}, while for protons and neutrons we use the equation
\begin{equation}\label{spec_heat}
C_{V}^{q}=T\frac{d S^{q}}{dT}\,,
\end{equation}

\noindent where we have defined the entropy of each species $S^{q}$ as

\begin{equation}\label{Eqentropy}
S^{q}=\sum_{\alpha} (2j_{\alpha}+1)[f^{q}_{\alpha}\ln f^{q}_{\alpha}+(1-f^{q}_{\alpha})\ln (1-f^{q}_{\alpha})]\,.
\end{equation}

\begin{figure}
\begin{center}
\includegraphics[width=0.42\textwidth,angle=-90]{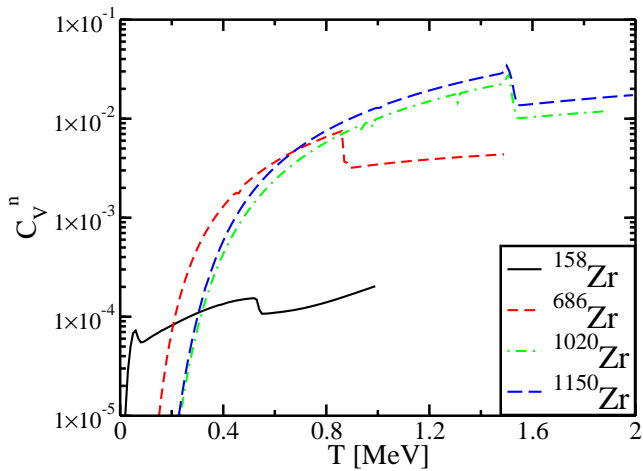}
\end{center}
\caption{(Color online) Neutron  specific heat of various WS cells presented in Fig.\ref{densCrust}.}
\label{180zr}
\end{figure}

\noindent  $f^{q}_{\alpha}=\left( 1+\exp \frac{E^{q}_{\alpha}}{T}\right)^{-1}$ is the Fermi distribution; $E^{q}_{\alpha}$ is the quasi-particle energy and $\alpha=\{nlj\}$ is a shorthand notation for the quantum number of the system.
In Fig.\ref{180zr}, we show the neutron specific heat $C_{V}^{n}$ as a function of the temperature for the different WS cells discussed in Fig.\ref{densCrust}.
For $^{686}$Zr,$^{1020}$Zr and $^{1150}$Zr we observe the presence of only one phase transition. The position of the critical temperature can be identified with Eq.\ref{Tcrit:bcs}. Beyond this value, the behavior of the specific heat can be described within the linear approximation for $C_{V}^{n}$. 
In fact, for these WS cells the condition $T \ll \varepsilon_{F}$ is satisfied. See discussion in Ref.~\cite{chamel10} for more details.

\begin{figure*}
\begin{center}
\includegraphics[width=0.43\textwidth,angle=-90]{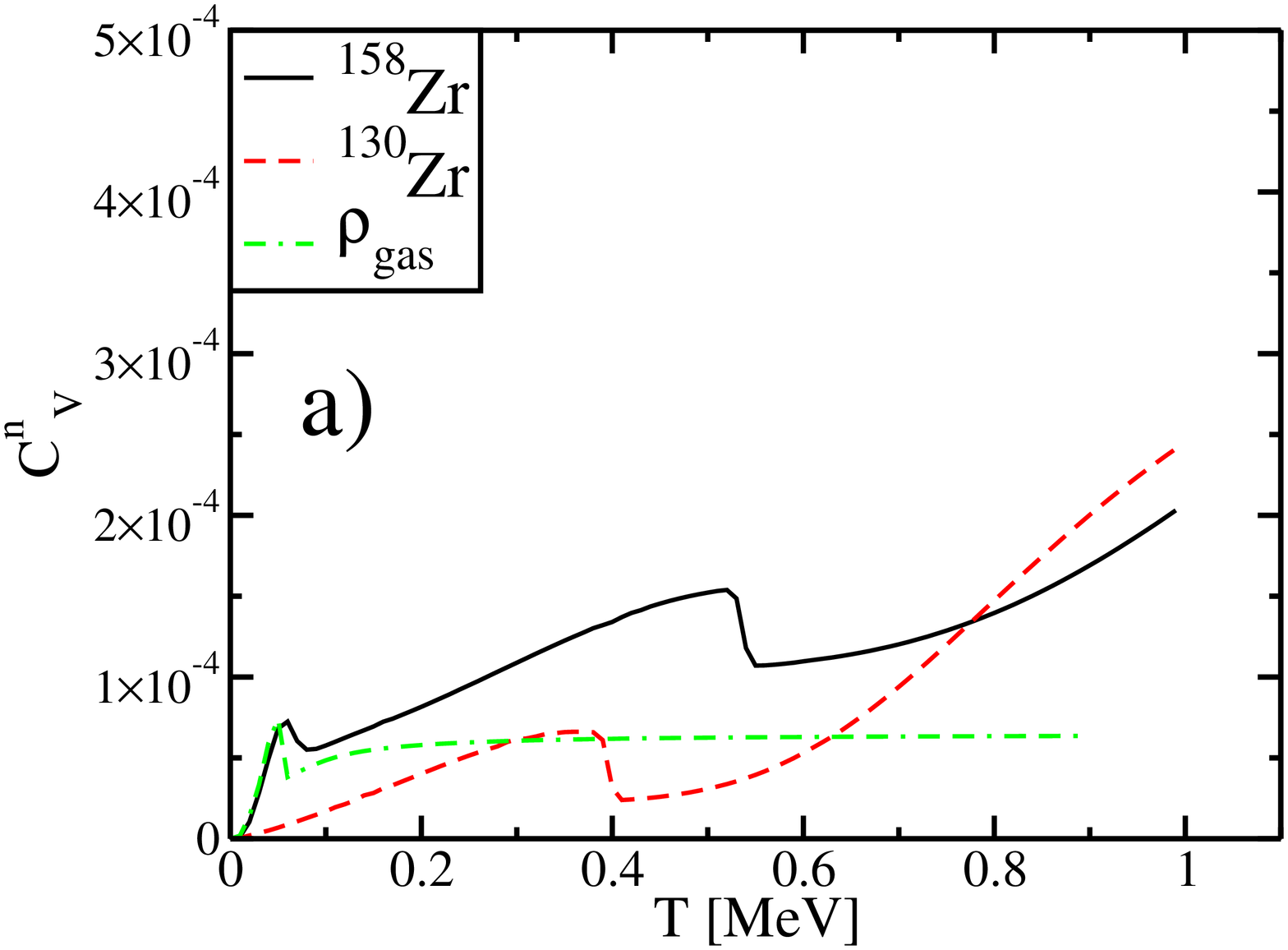}
\hspace{-3.2cm}
\includegraphics[width=0.43\textwidth,angle=-90]{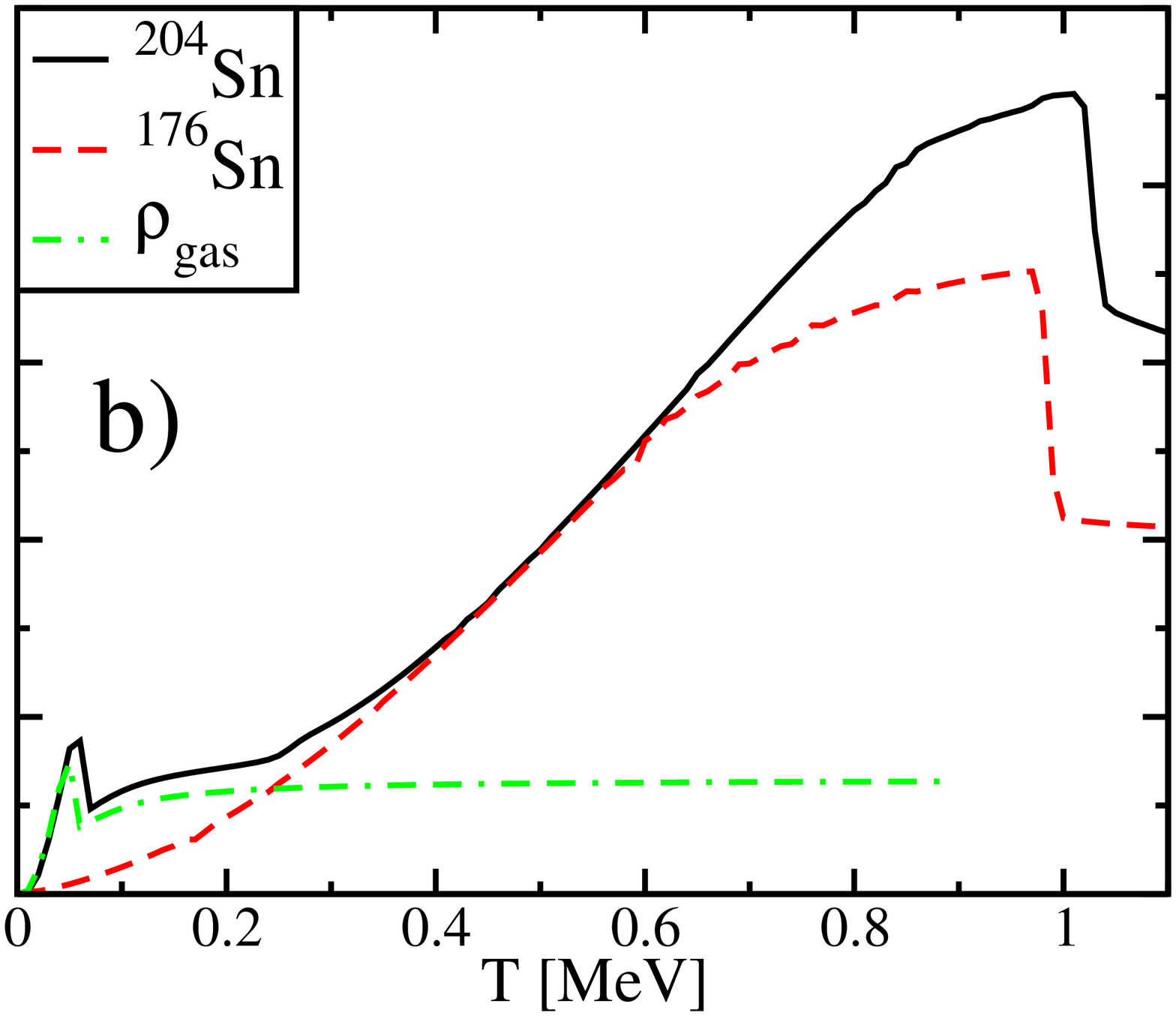}
\end{center}
\caption{(Color online) Neutron specific heat for different WS cells. With a dashed line, we represented the drip-line nucleus $^{130}$Zr and $^{176}$Sn respectively. The dashed-dotted line represents the specific heat for an external neutron gas with density $\rho_{gas}=5\times10^{-5}$ fm$^{-3}$.}
\label{180zr:serie}
\end{figure*}

For the WS cell $^{158}$Zr, which is at the interface between the outer and inner crust,  we can clearly identify two discontinuities in $C^{n}_{V}$ corresponding to two critical densities $T_{c,n}^{1,2}$
The first one $T_{c,1}^{n}=0.06$ MeV corresponds to the disappearance of pairing correlation in the external gas as it can also be seen from the behavior of the neutron pairing field $\Delta^{n}(R)$  in Fig.\ref{158pair}. At the second phase transition, $T_{c,2}^{n}=0.55$ MeV, the entire WS cell becomes non-superfluid. Beyond the critical density $T_{c,2}^{n}$, $C_{V}^{n}$ is not linear since in this case $T\approx \varepsilon_{F}$ and the behavior of the specific heat is more complex.For this particular configuration, the temperature at which the cluster evaporates is rather small ($\approx$3 MeV) MeV~\cite{mar03} and  already at $T\approx1$ MeV we observe important modification of the underlying shell structure. 
In Fig.~\ref{180zr:serie}, we compare  the neutron specific heat of $^{158}$Zr  with the specific heat of an homogenous gas of neutrons with density   $\rho_{gas}=5\times10^{-5}$ fm$^{-3}$ (panel a) and also the specific heat for $^{204}$Sn (panel b).
We notice that both WS cells have a first critical temperature at $T_{c,1}^{n}=60$ KeV. This temperature also coincides with the critical temperature of the uniform neutron gas. 
For  temperatures up to $T\approx0.2$ MeV, the properties of $C^{n}_{V}$ are essentially dominated by the external gas and thus showing a behavior which is roughly independent on the detailed composition of the crust. 
For higher values of the temperature the role of the cluster becomes more and more important. To this purpose we show on the same figure the specific heat calculated for the corresponding drip-line nucleus. 
For the $Sn$ case, we observe that the $C_{V}^{n}$ of $^{204}$Sn follows very closely the one of $^{176}$Sn up to $T\approx0.6$ MeV. Moreover the critical temperature for which the system becomes completely non-superfluid in both cases is rather close. This is not the case for $Zr$ isotopes, where none of these features is observed.


\section{Conclusions}\label{sec:conclusion}

We have studied the pairing properties of Wigner-Seitz cells at finite temperature by solving FT-HFB equations using the BSk21 functional. This functionals have been developed to reproduce with high accuracy both ground state properties, as masses and radii, of all know nuclei, but also some important properties of infinite nuclear matter~\cite{cha08b,duc12,pea14,gor09}.

We have devoted particular attention to the description of thermal properties of  low-density WS cells showing that the thermal effects  change their superfluid properties. Since the minimization procedure used to describe the chemical composition of the crust predicts several local minima which differers of few KeV per particle~\cite{duc12}, we could expect that the appearance (disappearance) of pairing correlations at finite temperature could play a non-negligible role.
Taking two representative WS cells, namely $^{158}Zr$ and $^{204}$Sn, we have also analyzed the properties of their neutron specific heat,  showing that in the low temperature regime, its behavior is almost independent on the nuclear cluster being mainly dominated by  the thermal properties of the external neutron gas. At higher densities the role of the cluster becomes more important.
An interesting analysis on the correlations between cluster and external neutron gas in WS cells has been performed in Ref.~\cite{pap13}. In that case the authors limited themselves to the zero-temperature case. It would be thus interesting to perform such kind of study for the finite temperature case. 
Recently several groups have demonstrated a lot of interest on the study of thermal properties of the inner crust in the low density region at the interface between outer and inner crust and at low-temperature.
The presence of pairing correlations strongly suppress the specific heat compared to the non-superfluid case. In this scenario, the heat capacity induced by the exchange of low-lying vibrational states ~\cite{khan05,sed10,bar10,cham13,pas14D,Martin14}  would play an important role. The coupling of lattice phonons with single-particle degrees of freedom also alters the pairing correlations of the WS cell~\cite{vigezzi2005pairing,bro07}. 
In the future it will be thus mandatory to treat all these different aspects on equal level to have a realistic description of the behavior of the low-density region of the inner crust of a neutron star. 


\section*{Acknowledgments}

Partial support comes from ÒNewCompStarÓ, COST Action MP1304. The author also thanks N. Chamel, S. Goriely and J. M. Pearson for useful and fruitful discussions which motivated the present article.


%
%
%
%
%
%

\bibliography{biblio}

\end{document}